\documentclass[11pt,fleqn,a4paper]{article} 
\usepackage{mathtools,amssymb,amsthm}
\usepackage{bm} 
\usepackage{physics} 
\usepackage{ascmac}
\usepackage{fancybox} 
\usepackage[bottom]{footmisc} 
\usepackage{graphicx}
\usepackage{here}
\usepackage{subcaption}
\usepackage{indentfirst}
\usepackage{tabularx}
\usepackage{multirow}
\usepackage{setspace} 
\usepackage{hyperref}
\usepackage{natbib}
\usepackage{tikz}
\usepackage{ifthen}
\theoremstyle{definition}
\newtheorem{dfn}{Definition}
\usepackage{enumerate}

\allowdisplaybreaks
\begin{document}
\title{The effect of minimum wages on employment in the presence of productivity fluctuations
\thanks{I am grateful to Gaston Chaumont for answering a wide range of questions and to Shinichi Nishiyama and Naoto Jinji for providing valuable comments and suggestions.}}
\author{Asahi Sato\footnote{Fourth-year student, Faculty of Economics, Kyoto University (email: sato.asahi.27d@st.kyoto-u.ac.jp).}}
\date{\today}
\maketitle

\begin{abstract}
\noindent
Traditionally, the impact of minimum wages on employment has been studied, and it is generally believed to have a negative effect.
Yet, some recent studies have shown that the impact of minimum wages on employment can sometimes be positive.
In addition, certain recent proposals set a higher minimum wage than the wage earned by some high-productivity workers.
However, the impact of minimum wages on employment has been primarily studied on low-skilled workers, whereas there is limited research on high-skilled workers.
To address this gap and examine the effects of minimum wages on high-productivity workers' employment, I construct a macroeconomic model incorporating productivity fluctuations, incomplete markets, directed search, and on-the-job search
and compare the steady-state distributions between the baseline model and the model with a minimum wage.
As a result, binding minimum wages increase the unemployment rate of both low and high-productivity workers.
\\ 
\\
\noindent\emph{JEL} Classification Numbers: E21, E24, J60. 
\end{abstract}

\setstretch{1.5}
\section{Introduction}
In recent years, the issue of raising the minimum wage has been a topic of extensive discussion.
Introducing or raising minimum wages has various economic effects.
In particular, the effects on employment have long been a subject of traditional research and
many empirical studies show that minimum wages have adverse impacts on employment \citep{neumarkniza2007minimum,eckstein2011job,sabia2012effects}.
Yet, some recent studies are critical of the adverse effects of minimum wages on employment \citep{cengiz2019effect,dustmann2022reallocation,engbom2022earnings}.
Thus, the effects of minimum wages on employment remain a subject of ongoing debate.
Though the effects of minimum wages are unclear, the increase in minimum wages has been steadily proceeding.
For example, as pointed out in \citet{hurst2022distributional}, in the United States, a proposal to raise the minimum wage from \$7.25 to \$15 is being considered
and about 40\% of non-college-educated workers and 10\% of college-educated workers earn a wage lower than \$15.
This implies that minimum wages have direct effects on not only low-productivity workers but also high-productivity workers.
However, most research examining the employment effects of minimum wages focuses on teenagers, low-wage industries,
and workers with low education levels \citep{manning2021elusive}.
This implies that only a few studies focus on high-productivity workers.
\par
Motivated by this, I examine the effects of minimum wages on the employment of high-productivity workers by using a macroeconomic model that incorporates productivity fluctuations and incomplete markets
a la \citet{aiyagari1994uninsured}, directed search, and on-the-job search.
In constructing the model for this paper, I referred to several papers.
\citet{hasumi2024general} incorporate \citet{aiyagari1994uninsured} type heterogeneous households and assume that wages are determined by bargaining.
The main difference between their research and mine is the labor market friction incorporated.
I choose directed search as I consider it to be more realistic in terms of households' knowing wages before applications.
\citet{zhao2021effects} add minimum wages to the model used in \citet{tsuyuhara2016dynamic}, which incorporates directed search and on-the-job search.
This research is very similar to mine, but in this research, households are assumed not to save.
I think that this is unrealistic so the model I use allows for household savings.
\par
This paper makes two contributions.
One is incorporating productivity fluctuations a la \citet{aiyagari1994uninsured} into a model that adopts the block recursive equilibrium.
As far as I know, no other paper combines these two.
The other is demonstrating that minimum wages raise the unemployment rate of both low and high-productivity workers.
\par
The next section describes the model proposed in this paper. Section 3 presents a quantitative analysis. The final section concludes.
\section{Model}
The model used in this paper closely follows \citet{chaumont2022wealth}, so almost all of the settings are the same.
However, for simplicity, I omit the assumption that unemployment insurance is calculated based on wages from the former position.
The main addition to \citet{chaumont2022wealth} is \citet{aiyagari1994uninsured} type productivity fluctuations of workers.
\subsection{Environment of the model}
Time is discrete and lasts forever. There is a unit measure of risk averse workers.
All workers are ex ante identical.
Each worker has a period utility function $u(c)$,
with $u'(0) = \infty$, $u'(c) > 0$ and $u''(c) < 0$ for all $c\in\mathbb{R}_{++}$.  
Workers and firms have the same discount factor $\beta \in (0,1)$.
Workers' assets $a$ must satisfy $a \geq \underline{a}$.
In numerical calculation, I set $\underline{a}$ to -0.5 so it can be thought as borrowing constraint.
The net rate of return on assets $r$ is assumed to be exogeneously determined.
In each period, both unemployed and employed workers search with probability 1 and $\lambda_{e} \in (0,1)$, respectively.
\par
$\varepsilon \in \qty{e, u}$ indicates a worker's employment status, where $e$ means being employed and $u$ means being unemployed.
A worker's productivity is $\eta \in \qty{l^{l}, l^{h}}$, where $l^{l}$ denotes low-productivity and $l^{h}$ denotes high-productivity.
Let $\omega$ be a worker's wages. Then, if a worker is employed, $\omega$ is equal to the worker's wage $w$, and if a worker is unemployed, $\omega = 0$.
\par
Firms are risk neutral and its measure is determined by competitive entry.
When firms enter the market, they create jobs and hire a worker.
The production technology has constant returns to scale with repect to jobs,
and jobs are independently operated.
Under these settings, each jobs can be treated as an individual firm, so I refer to a job as a firm.
Depending on the worker's productivity $\eta$, jobs yield an output $\eta y$ is produced, with zero production cost.
Firms cannot make a counteroffer and prohibit a worker from changing jobs when the worker gets a better offer.
Also, existing matches are subject to the exogeneous job destruction with probability $\delta\in (0,1)$,
which is i.i.d. across all jobs and periods.
\par
The labor market is partitioned into submarkets which are indexed by the applicant's productivity $\eta$, the wage offer $w$, and the applicant's wealth $a$.
Submarkets' index determines their tishtness $\theta(\eta, w, a)$, which is defined as the ratio of vacant jobs to applicants.
When firms and workers consider which submarket to enter, they think the tightness is predetermined and their choice does not change the value.
Since directed search is incorporated in this model, workers who select submarket $(\eta, w, a)$ must be endowed with productivity $\eta$ and the wealth $a$.
As explained in \citet{chaumont2022wealth}, incorporating applicants' wealth as an argument of market tightness simplifies the calculation of equilibrium.
\par
In submarket with tightness $\theta$, workers find jobs with probability $p(\theta)$
and vacant jobs are filled with probability $q(\theta)$, where $p(\theta) = \theta q(\theta)$.
I assume the following assumptions: $p(\theta)$, $q(\theta) \in [0,1]$, $p'(\theta) > 0$, $q'(\theta) < 0$
and $p^{\prime\prime}(\theta) < 0$ for all $\theta \in \mathbb{R}_{+}$.
\par
Each period is divided into five stages: (\romannumeral1) production, (\romannumeral2) consumption and savings, (\romannumeral3) search and matching,
(\romannumeral4) exogenous separation, and (\romannumeral5) productivity fluctuations.
In stage (\romannumeral1), all existing matches produce, and workers gain wages. In stage (\romannumeral2), workers determine how much they consume and save.
In stage (\romannumeral3), both unemployed and employed workers search with probability 1 and $\lambda_{e}$, respectively.
In stage (\romannumeral4), existing matches are subject to exogenous destruction at rate $\delta \in (0,1)$ and employed workers become unemployed.
However, new matches formed in current period are assumed to be immune to that separation until next period.
In stage (\romannumeral5), each worker's productivity fluctuates according to the 2-state Markov chain which is explained later in section 3.1.
\par
Workers are characterized by their employment status, productivity, earnings, and assets.
$s \equiv (\varepsilon,\eta,\omega,a)\in S\equiv E\times L\times \bar{W}\times A$ represents a worker's individual state,
where $E\equiv \qty{e,u}$, $L\equiv \qty{l^l,l^h}$, $\bar{W}\equiv [\underline{w},\bar{w}] \cup \qty{0}$, and $A\equiv [\underline{a},\bar{a}]$.
Define the Borel sets as $\mathcal{E}$ corresponding to $E$, $\mathcal{L}$ corresponding to $L$, $\mathcal{W}$ corresponding to $\bar{W}$, and $\mathcal{A}$ corresponding to $A$.
Let $\mathcal{S}=\mathcal{E}\times\mathcal{L}\times\mathcal{W}\times\mathcal{A}$ be the Borel set corresponds to the individual state space $S$.
\par
Probability measure $\psi \colon \mathcal{S}\to [0,1]$ gives a distribution of workers over individual states and expresses the aggregate state of the economy. 
Let $\Psi (S,\mathcal{S})$ be the space of distribution functions on the measurable space $(S,\mathcal{S})$.
The law of motion for the state of the economy is given by the operator $\mathcal{T} \colon \Psi (S,\mathcal{S})\to \Psi (S,\mathcal{S})$.
\subsection{Decisions over consumption and savings}
Consider a worker's state is $(\varepsilon, \eta, \omega, a)$.
As in \citet{chaumont2022wealth}, the worker's value function is defined as follows:
\begin{equation}
  \begin{aligned}
    & V_{\varepsilon}(\eta,\omega,a) = \max _{(c, \hat{a})}\left[u(c)+\beta R_{\varepsilon}(\eta,\omega,\hat{a})\right] \\
    & \text { s.t. } c+\frac{\hat{a}}{1+r}=y_{\min} + \omega +a \quad \text { and } \quad \hat{a} \geq \underline{a}, \label{vf}
  \end{aligned}
\end{equation}
where $\hat{a}$ is the assets at the beginning of the next period\footnote{The hat is used to distinguish between current and next period.} and $R_{\varepsilon}(\eta,\omega,\hat{a})$ is also a value function about the return of search defined later.
The $y_{\text{min}}$ is a small amount of home production to prevent $c \leq 0$.
Let $c_{\varepsilon}(\eta,\omega,a)$ and $\hat{a}_{\varepsilon}(\eta,\omega,a)$ be the policy function for optimal consumption and optimal savings, respectively.
\subsection{Unemployed workers' search}
Consider a low-productivity unemployed worker.
Let $R_{u}(\eta,0,\hat{a})$ represents the value attained by the optimal search.
When the worker selects submarket $(l^l, \hat{w}, \hat{a})$, the worker gets a job with probability $p(\theta(l^l, \hat{w}, \hat{a}))$
and attain the value $T_{ll} V_e(l^l, \hat{w}, \hat{a}) + T_{lh} V_e(l^h, \hat{w}, \hat{a})$,
where $T_{ll}$ and $T_{lh}$ are the productivity transition probabilities.
The subscript $_{ll}$ denotes transition from low to low productivity and $_{lh}$ means from low to high productivity.
The same goes for $_{hl}$ and $_{hh}$ as well.
On the other hand, the worker remains unemployed with probability $1-p(\theta(l^l, \hat{w}, \hat{a}))$ and attain the value $T_{ll} V_u(l^l, 0, \hat{a}) + T_{lh} V_u(l^h, 0, \hat{a})$.
Except for the transition probability and the probabilities of being hired, the same can be said for high-productivity workers. Then,
\begin{equation}
  \begin{aligned}
    R_u(l^l, 0, \hat{a}) = \max _{\hat{w}} & \left\{p(\theta(l^l, \hat{w}, \hat{a})) (T_{ll} V_e(l^l, \hat{w}, \hat{a}) + T_{lh} V_e(l^h, \hat{w}, \hat{a})) \right. \\
    & \left.+(1-p(\theta(l^l, \hat{w}, \hat{a}))) (T_{ll} V_u(l^l, 0, \hat{a}) + T_{lh} V_u(l^h, 0, \hat{a}))\right\}
  \end{aligned}
\end{equation}
and
\begin{equation}
  \begin{aligned}
    R_u(l^h, 0, \hat{a}) = \max _{\hat{w}} & \left\{p(\theta(l^h, \hat{w}, \hat{a})) (T_{hl} V_e(l^l, \hat{w}, \hat{a}) + T_{hh} V_e(l^h, \hat{w}, \hat{a}))\right. \\
    & \left.+(1-p(\theta(l^h, \hat{w}, \hat{a}))) (T_{hl} V_u(l^l, 0, \hat{a}) + T_{hh} V_u(l^h, 0, \hat{a}))\right\}.
  \end{aligned}
\end{equation}
Since consumption and savings stage precedes search and matching stage, optimal search policy is a function of $(\eta,0,\hat{a})$.
$\hat{a}$ is a function of $(\eta,0,a)$.
Thus, optimal search policy $\hat{w}_{u}$ is also a function of $(\eta,0,a)$.
Moreover, since both $\hat{w}$ and $\hat{a}$ are functions of $(\eta,0,a)$, the tightness of the optimal submarket for the unemployed worker $\hat{\theta}_{u}$ is also a function of $(\eta,0,a)$.
\subsection{Employed workers' search}
Consider a low-productivity employed worker with wage $w$ and wealth $\hat{a}$. 
The worker is allowed to search with probability $\lambda_e$.
When the worker selects the submarket $(l^l, \hat{w}, \hat{a})$, the worker gets a job with probability $p(\theta(l^l, \hat{w}, \hat{a}))$ and gain the value $T_{ll} V_e(l^l, \hat{w}, \hat{a}) + T_{lh} V_e(l^h, \hat{w}, \hat{a})$.
Even if the worker fails to transit to other jobs, the worker does not immediately become unemployed.
The exogeneous job destruction hits the worker with probability $\delta$.
If the worker avoid it, the worker keep the current position, so the worker gets the value $T_{ll}V_{e}(l^l,w,\hat{a}) + T_{lh}V_{e}(l^h,w,\hat{a})$.
If not, the worker gets the value $T_{ll}V_{u}(l^l,0,\hat{a}) + T_{lh}V_{u}(l^h,0,\hat{a})$.
Except for the transition probability and the probabilities of being hired, the same can be said for high-productivity workers.
Thus,
\begin{equation}
  \begin{aligned}
    R_e(l^l, w, \hat{a}) = \max _{\hat{w}} & \left\{\lambda_{e}p(\theta(l^l, \hat{w}, \hat{a})) (T_{ll} V_e(l^l, \hat{w}, \hat{a}) + T_{lh} V_e(l^h, \hat{w}, \hat{a}))\right. \\
    & +(1-\lambda_{e}p(\theta(l^l, \hat{w}, \hat{a})))T_{ll} [(1-\delta)V_{e}(l^l,w,\hat{a}) + \delta V_u(l^l, 0, \hat{a})] \\
    & \left.+(1-\lambda_{e}p(\theta(l^l, \hat{w}, \hat{a}))) T_{lh} [(1-\delta)V_{e}(l^h,w,\hat{a})  + \delta V_u(l^h, 0, \hat{a})]\right\}
  \end{aligned}
\end{equation}
and
\begin{equation}
  \begin{aligned}
    R_e(l^h, w, \hat{a}) = \max _{\hat{w}} & \left\{\lambda_{e}p(\theta(l^h, \hat{w}, \hat{a})) (T_{hl} V_e(l^l, \hat{w}, \hat{a}) + T_{hh} V_e(l^h, \hat{w}, \hat{a}) )\right. \\
    & +(1-\lambda_{e}p(\theta(l^h, \hat{w}, \hat{a})))T_{hl} [(1-\delta)V_{e}(l^l,w,\hat{a}) + \delta V_u(l^l, 0, \hat{a})] \\
    & \left.+(1-\lambda_{e}p(\theta(l^h, \hat{w}, \hat{a}))) T_{hh} [(1-\delta)V_{e}(l^h,w,\hat{a})  + \delta V_u(l^h, 0, \hat{a})]\right\}.
  \end{aligned}
\end{equation}
Let $\hat{w}_{e}$ and $\hat{\theta}_e$ be the optimal search policy and the tightness of the optimal submarket for the employed worker, respectively.
For the same reason explained in the above subsection, both of them are a function of $(\eta,w,a)$.
\subsection{Firms and market tightness}
The value of filled job consists of current profits and future value of the job.
If a match is dissolved, the future value is 0.
Exogeneous and endogenous separation occurs with probability $\delta$ and $\lambda_{e}p(\hat{\theta}_{e}(l^l,w,a))$, respectively.
Thus, the future value takes $\beta \qty{T_{ll}J(l^l,w,\hat{a}_{e}(l^l,w,a)) + T_{lh}J(l^h,w,\hat{a}_{e}(l^l,w,a))}$ with probability $(1-\delta)[1-\lambda_{e}p(\hat{\theta}_{e}(l^l,w,a))]$.
Except for the output and the matching probabilities, the same can be said for high-productivity workers.
Then, the value of filled jobs is:
\begin{equation}
  \label{jl}
  J(l^l,w,a) = l^{l}y - w + (1-\delta)[1-\lambda_{e}p(\hat{\theta}_{e}(l^l,w,a))]\beta \qty{T_{ll}J(l^l,w,\hat{a}_{e}(l^l,w,a)) + T_{lh}J(l^h,w,\hat{a}_{e}(l^l,w,a))}
\end{equation}
and
\begin{equation}
  \label{jh}
  J(l^h,w,a) = l^{h}y - w +  (1-\delta)[1-\lambda_{e}p(\hat{\theta}_{e}(l^h,w,a))]\beta \qty{T_{hl}J(l^l,w,\hat{a}_{e}(l^h,w,a)) + T_{hh}J(l^h,w,\hat{a}_{e}(l^h,w,a))}.
\end{equation}
Competitive entry of vacancies determines each submarket tightness.
Consider submarket $(l^l, \hat{w}, \hat{a})$.
If there is potential surplus, that is, the future value of the job is larger than cost of posting a vacancy ($\beta\qty{T_{ll} J\qty(l^l,\hat{w},\hat{a}) + T_{lh} J\qty(l^h,\hat{w},\hat{a})} \geq k$),
by free entry, vacancies are created until the future value of the job and the cost of posting a vacancy take the same value.
Thus, the following equation holds:
\begin{equation}
  0 = -k + \beta q(\theta(l^l,\hat{w},\hat{a}))\qty{T_{ll}J(l^l,\hat{w},\hat{a})+T_{lh}J(l^h,\hat{w},\hat{a})}.
\end{equation}
If there is no potential surplus, that is, if $k > \beta\qty{T_{ll} J\qty(l^l,\hat{w},\hat{a}) + T_{lh} J\qty(l^h,\hat{w},\hat{a})}$,
no vacancies are posted.
Then, $\theta\qty(l^l,\hat{w},\hat{a})=0$.
From these two, the following equation can be derived:
\begin{equation}
  \label{thetal}
  \theta\qty(l^l,\hat{w},\hat{a})=
      \begin{cases}
        q^{-1}\qty(\frac{k}{\beta\qty{T_{ll} J\qty(l^l,\hat{w},\hat{a}) + T_{lh} J\qty(l^h,\hat{w},\hat{a})}})   &   \text{if}\, \text{$\beta\qty{T_{ll} J\qty(l^l,\hat{w},\hat{a}) + T_{lh} J\qty(l^h,\hat{w},\hat{a})}$} \geq k  \\
        0        &   \text{otherwise}.
      \end{cases}
\end{equation}
Same goes for $\theta\qty(l^h,\hat{w},\hat{a})$. Thus,
\begin{equation}
  \label{thetah}
  \theta\qty(l^h,\hat{w},\hat{a})=
      \begin{cases}
        q^{-1}\qty(\frac{k}{\beta\qty{T_{hl} J\qty(l^l,\hat{w},\hat{a}) + T_{hh} J\qty(l^h,\hat{w},\hat{a})}})   &   \text{if}\, \text{$\beta\qty{T_{hl} J\qty(l^l,\hat{w},\hat{a}) + T_{hh} J\qty(l^h,\hat{w},\hat{a})}$} \geq k  \\
        0        &   \text{otherwise}.
      \end{cases}
\end{equation}
\subsection{Equilibrium}
\begin{dfn}
  An equilibrium consists of the value function of workers $V_{\varepsilon}$ (for $\varepsilon = e,u$), the firm value function $J$,
  policy functions ($c_{\varepsilon},\hat{a}_{\varepsilon}$) and ($\hat{w}_{u},\hat{w}_{e}$), and the transition of the aggregate state $\mathcal{T}$,
  such that: (\romannumeral1) The value functions of workers $V_{\varepsilon}\colon L\times \bar{W}\times A \to \mathbb{R}$ satisfies \eqref{vf}, optimal decisions of consumption and savings
  yield the policy functions ($c_{\varepsilon},\hat{a}_{\varepsilon}$), and optimal search decisions yield the policy functions ($\hat{w}_{u},\hat{w}_{e}$);
  (\romannumeral2) the firm value function $J\colon L\times W\times A \to \mathbb{R}$ satisfies \eqref{jl} and \eqref{jh};
  (\romannumeral3) the tightness function $\theta$ satisfies \eqref{thetal} and \eqref{thetah} for all $(l^l,\hat{w},\hat{a})$, $(l^h,\hat{w},\hat{a}) \in L\times W\times A$;
  and (\romannumeral4) the aggregate state transition $\mathcal{T}\colon \Psi(S,\mathcal{S})\to \Psi(S,\mathcal{S})$ is consistent with the policy functions.
\end{dfn}
This definition of equilibrium closely follows that in \citet{chaumont2022wealth}.
Since I add productivity fluctuations, $L$ is added to the domain of value function $V_{\varepsilon}$ and $J$.
Also, condition (\romannumeral3) is changed a little bit.
For the properties of this equilibrium type, see section 3 of \citet{chaumont2022wealth}.
\section{Quantitative analysis}
I choose parameters to reproduce the unemployment rate and labor market transition rates in Japan for 2023.
\subsection{Calibration}
The utility function have the following forms:
\begin{equation}
  u(c)=\frac{c^{1-\sigma}}{1-\sigma}.
\end{equation}
Following \citet{menzio2010block}, the matching probabilities have the following forms:
\begin{equation}
  p(\theta)=\theta(1+\theta^{\gamma})^{-\frac{1}{\gamma}},\quad q(\theta)=\frac{p(\theta)}{\theta}.
\end{equation}
\par
I calibrate the model to the monthly frequency. Normalize $y$ = 1.
\begin{table}[tp]							
  \caption{Main parameter values.}
  \label{T:Parameters}
  \begin{center}\begin{small}
  \begin{tabularx}{\textwidth}{@{\extracolsep{\fill}}lXlX}
  \hline\hline\vphantom{\rule{0pt}{12pt}}%
  Parameters & Descriptions & Values &	Comments / Targets \\
  \hline\vphantom{\rule{0pt}{12pt}}%
   $\beta$ & discount factor & 0.996	& when raised to the power of three, this is $\fallingdotseq$ 0.987 used in \citet{menzio2010block} \\
   $r$ & the net rate of return on assets & 0.327\% & annual interest rate = 4.0\% \\
   $\underline{a}$ & borrowing limit & -5.0 & no specific reason \\
   $\delta$ & exogeneous separation rate & 0.633\% & calculated based on the data from Labour Force Survey \\
   $\sigma$ & the risk aversion parameter & 2.0 & often used in macro calibration \\
   $\gamma$ & the matching technology parameter & 0.5 & following \citet{menzio2010block} \\
   $\lambda_{e}$ & the probability of having the opportunity to search & 0.0346 & job-to-job transition rate = 0.41\% \\
   $k$ & the cost of posting a job & 0.77 & unemployment rate = 2.6\% \\
  \hline\hline
  \end{tabularx}
  \end{small}\end{center}
\end{table}
Table\ref{T:Parameters} lists the parameters and calibration targets.
The monthly interest rate is $r$ = 0.327\%, or 4.0\% annually. I set $\beta$ to 0.996.
When raised to the power of three, this is $\fallingdotseq$ 0.987, which is used in \citet{menzio2010block}\footnote{In the paper, the period is set to be one quarter.}.
I set $\underline{a}$ to -5.0.
The risk aversion parameter is set to $\sigma$ = 2, which is commonly used in the literature.
\par
I set the exogenous separation rate $\delta$ = 0.00633 $\fallingdotseq$ 430000/(430000+67470000).
The value 430000 is the number of involuntary job separation and the value 67470000 is the number of employees.
Both values are from 2023 Yearly Average Results of Labour Force Survey \citep{labour_force_survey2023}.
For an employed worker, the probability of having the opportunity to search $\lambda_{e}$ is chosen to yield a monthly job-to-job transition rate of 0.41\% $\fallingdotseq$ (3280000/12)/67470000.
The value 3280000 is the number of job changers during 2023 taken from 2023 Yearly Average Results (Detailed Tabulation) of Labour Force Survey \citep{labour_force_survey_d2023}.
Following \citet{menzio2010block}, the matching technology parameter $\gamma$ is set to be 0.5.
Following \citet{chaumont2022wealth}, unearned income $y_{min}$ is set to be 0.05.
The vacancy cost $k$ is set to be 0.77 to match an unemployment rate of 2.6\% $\fallingdotseq$ 1780000/(1780000+67470000).
The value 1780000 is the number of unemployed person taken from 2023 Yearly Average Results of Labour Force Survey \citep{labour_force_survey2023}.
\par
I assume that labor productivity follow an AR(1) process as follows:
\begin{equation}
  \log (l_{t}) = \rho \log (l_{t-1}) + \epsilon _{t},\quad \epsilon_{t}\sim N(0,\sigma_{2}^{2}).
\end{equation}
I discretize it with $\rho=0.99$ and $\sigma_{2}=0.05$ into a 2-state Markov chain by using the method of \citet{rouwenhorst1995asset}.
As a result, $l^l$ is set to be $0.7016$ and $l^h$ is set to be $1.4254$.
Also, $T_{ll}$ is set to be 0.995, $T_{lh}$ is set to be 0.005, $T_{hl}$ is set to be 0.005, and $T_{hh}$ is set to be 0.995.
\subsection{Solution method and stationary distribution}
I define the grids for the spaces of wealth and wages and set the number of grid points to 141 and 20, respectively.
Wealth is set to be in $[\underline{a},\bar{a}] = [-5.0,700]$ and wages are set to be in $[\underline{w},\bar{w}] = [0,1.4254]$.
To solve the model, I use the following algorithm:
\begin{enumerate}
\item Guess initial value functions of workers and firms.
\item Given the value functions of firms, solve for the market tightness using \eqref{thetal} and \eqref{thetah}.
\item Given the tightness function, solve the worker's optimization problem and compute optimal savings, consumption and job search decisions.
In this step, I use Value Function Iteration, that is, this step is iterated until resulting value functions converge.
\item Given worker's policy functions computed in the previous step, calculate separation rates of employed workers and update the value function of the firm in each submarket.
Until convergence, I iterate this step.
\item Check the difference between initial value functions of firm and previously computed ones.
If the difference is sufficiently small, end this algorithm. If not, using calculated value functions instead of initial guess, repeat steps 2 to 4 above.
\end{enumerate}
\par
After solving the model, I derive the transition probability matrix for household state variables using policy functions,
and also, I obtain stationary distribution by use of this transition probability matrix.
\subsection{Results in the baseline model}
\begin{table}[tp]
  \caption{Indicator values in the baseline model.}
  \label{T:Ivb}
  \begin{center}\begin{small}
  \begin{tabularx}{\textwidth}{@{\extracolsep{\fill}}ll}
  \hline\hline\vphantom{\rule{0pt}{12pt}}%
  Indicator & Values \\
  \hline\vphantom{\rule{0pt}{12pt}}%
  Unemployment rate of low-productivity workers & 1.396\%  \\
  Unemployment rate of high-productivity workers & 1.178\%  \\
  Job-to-job transition rate & 0.406\% \\
  \hline\hline
  \end{tabularx}
  \end{small}\end{center}
\end{table}
Table \ref{T:Ivb} shows the main resulting indicator values in the baseline model.
The unemployment rate for low-productivity workers is 1.396\%, which is higher than that for high-productivity workers.
The job-to-job transition rate in the baseline model is 0.406\%.
\par
Figure 1 depicts a worker's optimal search decisions without minimum wages as a function of current wealth.
In the legend, I use w1, w2, \dots, w20, which represent the 1st grid point, 2nd grid point, \dots, 20th grid point, respectively.
As shown in this figure, the more wealth workers have, the higher wage workers target.
The submarket tightness $\theta$ is decreasing in wage so that the matching probability is also decreasing in wage.
Because of this, workers with enough wealth tend to run the risk of losing a job by applying for a higher-paying job.
When comparing high-productivity workers and low-productivity workers, high-productivity workers are predisposed to apply for a job with a higher wage.
When other arguments ($\hat{w}$ and $\hat{a}$) are the same, $\theta(l^h,\hat{w},\hat{a})\geq \theta(l^l,\hat{w},\hat{a})$ is always valid
because the value of a job with a high-productivity worker is higher than that with a low-productivity worker.
That is why, a high-productivity worker apply for a higher-paying job.
From panels (b) and (c), we can see that as the wage of the current job increases, workers target higher wages.
This is because there is no reason for an employed worker to apply for a job with lower wages than those of the worker's current job.
In panel (b), it seems that the lines indexed by w12, w13, \dots, w19 are not displayed.
Likewise, in panel (c), the lines indexed by w18 and w19 do not seem to be displayed.
They are hidden under the line ``w20''.
This phenomenon does not stem from a mistake in my code and is not an issue at all. The vacancy post cost $k$ is so high relative to output that as shown in Figure 6, the market tightnes falls to zero quickly.
Because of this, the market tightness of a job for which a worker with highly-paid job want to apply is 0.
Then, for employed workers with well-paying job, it does not matter which wages they target.
\par
Figure 2 depicts the difference between next period's wealth and current wealth without minimum wages.
From panel (a), it can be understood that high-productivity unemployed workers tend to save less wealth.
They will earn more relative to low-productivity workers with high probabilities so that they consume more.
Panels (b) and (c) show that there are no significant differences in the saving policy depending on the productivity of an employed worker.
Moreover, from this figure, we can see that saving is a decreasing function of wealth at the beginning of the period.
If the amount of wealth at the beginning of the period is small, it is very important for workers to save more because they have more chances of getting hired.
However, if a worker has a sufficient amount of wealth at the beginning of the period, it is optimal for the worker to save less and gain instant utility at the cost of wealth at the beginning of the next period.
\par
From Figure 3, we can see that the more wealth workers have, the more they consume.
We can see that from panel (a), high-productivity unemployed workers consume more.
Panels (b) and (c) show that as the wage of the current job increases, workers consume more.
It is not surprising that consumption policy is an increasing function of both wealth and productivity.
\par
Figure 4 depicts the stationary distribution without minimum wages.
Panel (a) shows that in the equilibrium, earnings tend to be concentrated at the higher end.
Ten bars\footnote{Note that although it is barely visible, there is the second bar from the left to the right of the number 0.5.} appear in Panel (a).
The third bar from the left represents the proportion of employed workers with "w10" wage ($\bar{w}\times (10-1)/(20-1) \fallingdotseq 0.6752$).
As stated in the section 3.4, in testing the effects of minimum wages, I set $\underline{w}$ to 0.1 and 0.7.
The former expresses non-binding minimum wages, while the latter expresses binding minimum wages.
Panel (b) shows that in the equilibrium, the distribution of wealth is characterized by a long tail on the right side.
\par
In Figure 5, the firm value of a filled job is expressed as a function of wealth or wage.
Because there are so many grid points in the space of wealth, in panels (c) and (d), I pick up and display lines indexed by 1st, 12th, and 141st grid point.
When comparing panels (a) and (b) or (c) and (d), it can be seen that the value of a firm with a high-productivity worker is higher.
We can see that as wages increase, the value of the job decreases. Furthermore, the job value is an increasing function of wealth.
\par
Figure 6 depicts market tightness as a function of wealth or wage.
Because there are so many grid points in the space of wealth, in panels (c) and (d), I pick up and display lines indexed by 1st, 12th, and 141st grid point.
When comparing panels (a) and (b) or (c) and (d), it can be understood that the market tightness for a high-productivity worker is higher.
We can see that as wages values increase, the market tightness decreases. Furthermore, the market tightness is an increasing function of wealth.
Note that because the market tightness $\theta$ is the ratio of vacancies to applicants, the higher the market tightness is, the more advantageous for workers.
Why is the value of job and market tightness an increasing function of wealth?
As shown in Figure 1, more wealth causes a worker to apply for a job with higher wages.
This means that it is difficult for employed workers with significant wealth to change jobs so that the job conditioned by much wealth is less likely to expire.
Then, the value of job is an increasing function of wealth.
Because the market tightness rises as the job value increases, the market tightnes is also an increasing function of wealth.
\begin{figure}[H]
  \centering
  \begin{subfigure}[t]{\textwidth}
      \centering
      \includegraphics[width=0.66\textwidth]{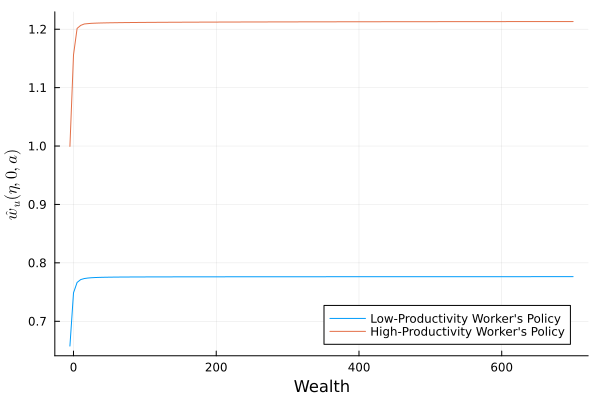}
      \caption{Search Policy for Unemployed Workers}
  \end{subfigure}
  \vspace{1em} 
  \begin{subfigure}[t]{\textwidth}
      \centering
      \includegraphics[width=0.66\textwidth]{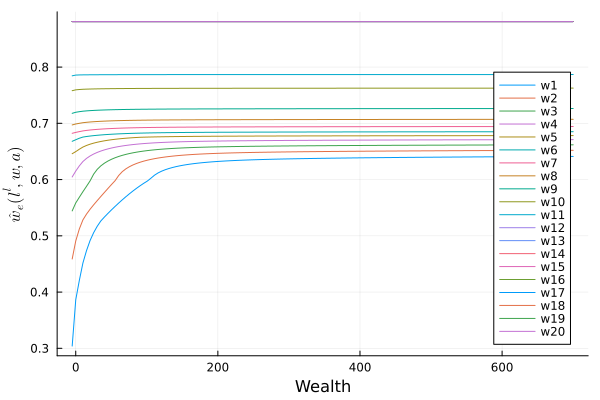}
      \caption{Search Policy for Low-Productivity Employed Workers}
  \end{subfigure}
  \vspace{1em} 
  \begin{subfigure}[t]{\textwidth}
      \centering
      \includegraphics[width=0.66\textwidth]{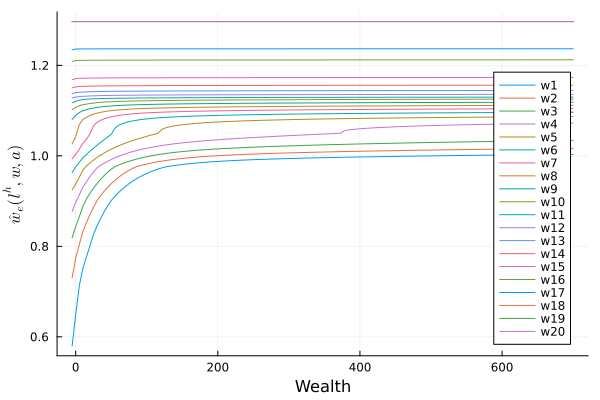}
      \caption{Search Policy for High-Productivity Employed Workers}
  \end{subfigure}
  \caption{Optimal search policy.}
\end{figure}
\begin{figure}[H]
  \centering
  \begin{subfigure}[t]{\textwidth}
      \centering
      \includegraphics[width=0.66\textwidth]{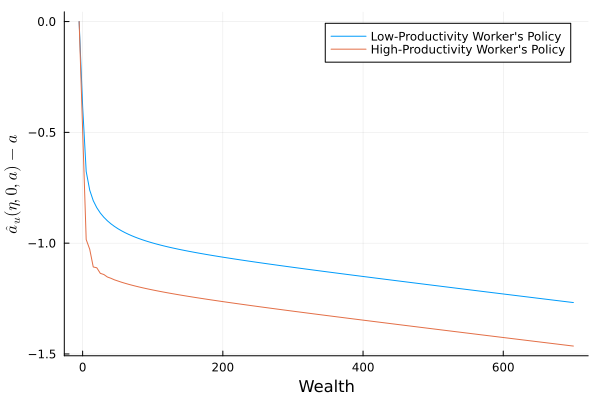}
      \caption{Saving Policy for Unemployed Workers}
  \end{subfigure}
  \vspace{1em} 
  \begin{subfigure}[t]{\textwidth}
      \centering
      \includegraphics[width=0.66\textwidth]{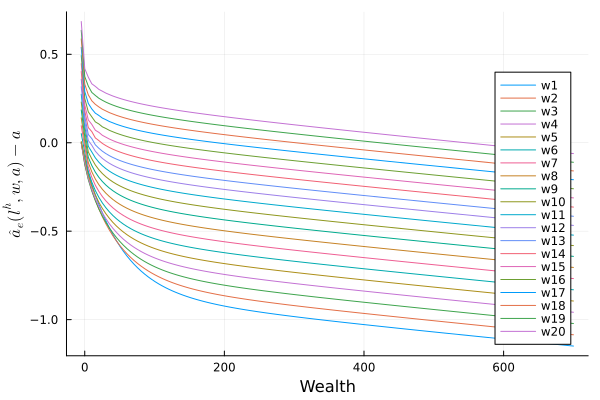}
      \caption{Saving Policy for Low-Productivity Employed Workers}
  \end{subfigure}
  \vspace{1em} 
  \begin{subfigure}[t]{\textwidth}
      \centering
      \includegraphics[width=0.66\textwidth]{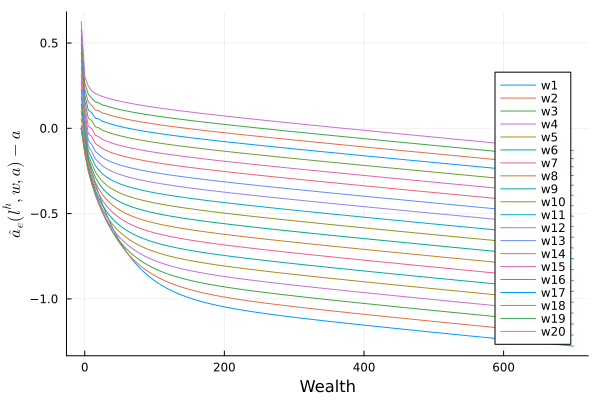}
      \caption{Saving Policy for High-Productivity Employed Workers}
  \end{subfigure}
  \caption{Optimal saving policy.}
\end{figure}
\begin{figure}[H]
  \centering
  \begin{subfigure}[t]{\textwidth}
      \centering
      \includegraphics[width=0.66\textwidth]{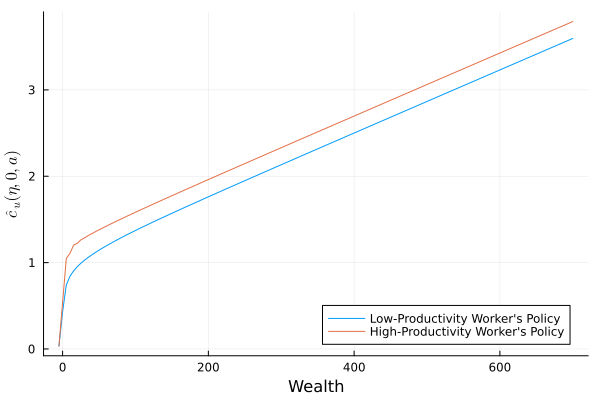}
      \caption{Consumption Policy for Unemployed Workers}
  \end{subfigure}
  \vspace{1em} 
  \begin{subfigure}[t]{\textwidth}
      \centering
      \includegraphics[width=0.66\textwidth]{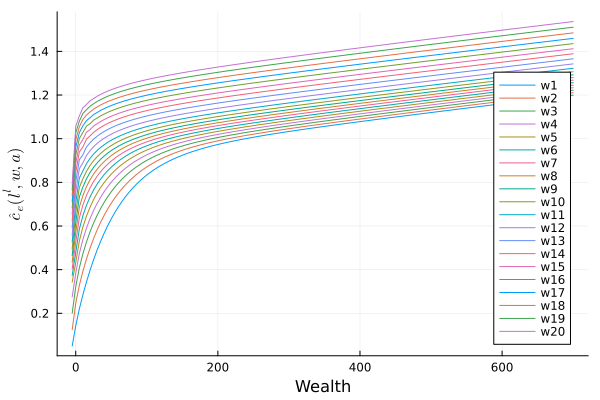}
      \caption{Consumption Policy for Low-Productivity Employed Workers}
  \end{subfigure}
  \vspace{\baselineskip} 
  \begin{subfigure}[t]{\textwidth}
      \centering
      \includegraphics[width=0.66\textwidth]{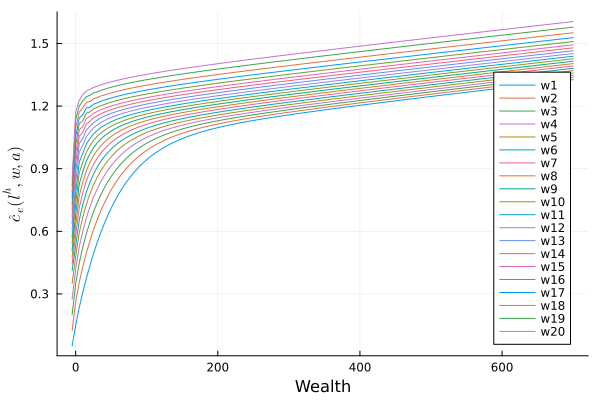}
      \caption{Consumption Policy for High-Productivity Employed Workers}
  \end{subfigure}
  \caption{Optimal consumption policy.}
\end{figure}
\begin{figure}[H]
  \centering
  \begin{subfigure}[t]{\textwidth}
      \centering
      \includegraphics[width=1.00\textwidth]{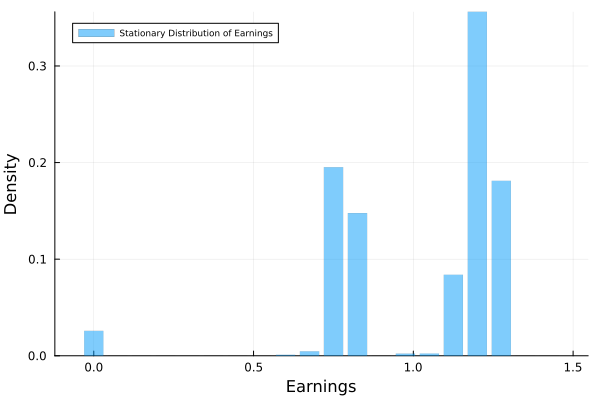}
      \caption{Stationary Distribution of Earnings}
  \end{subfigure}
  \vspace{1cm} 
  \begin{subfigure}[t]{\textwidth}
      \centering
      \includegraphics[width=1.00\textwidth]{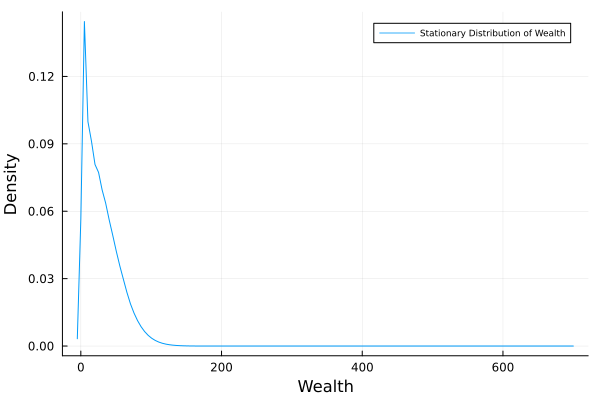}
      \caption{Stationary Distribution of Wealth}
  \end{subfigure}
  \caption{Stationary distribution.}
\end{figure}
\begin{figure}[H]
  \begin{tabular}{cc}
  \begin{minipage}[b]{0.45\textwidth}
  \centering
  \includegraphics[keepaspectratio, width=0.9\textwidth]{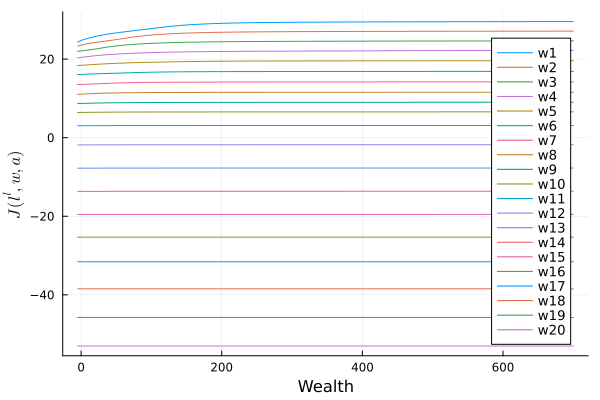}
  \subcaption{Firms Value with a Low-Productivity Worker and Wealth}
  \end{minipage}&
  \begin{minipage}[b]{0.45\textwidth}
  \centering
  \includegraphics[keepaspectratio, width=0.9\textwidth]{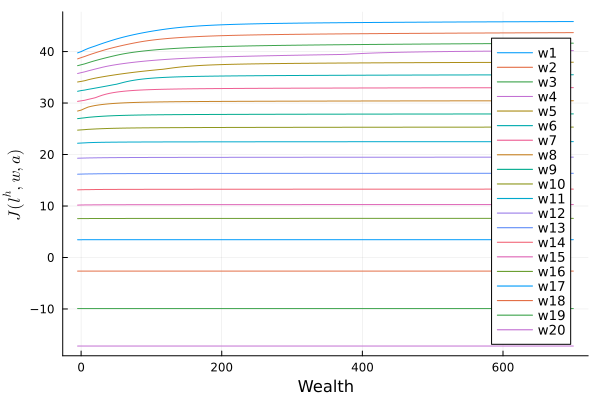}
  \subcaption{Firms Value with a High-Productivity Worker and Wealth}
  \end{minipage}\\
  \begin{minipage}[b]{0.45\textwidth}
  \centering
  \includegraphics[keepaspectratio, width=0.9\textwidth]{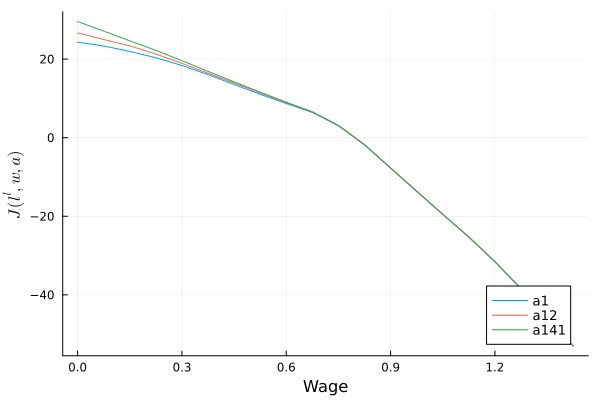}
  \subcaption{Firms Value with a Low-Productivity Worker and Wage}
  \end{minipage}&
  \begin{minipage}[b]{0.45\textwidth}
  \centering
  \includegraphics[keepaspectratio, width=0.9\textwidth]{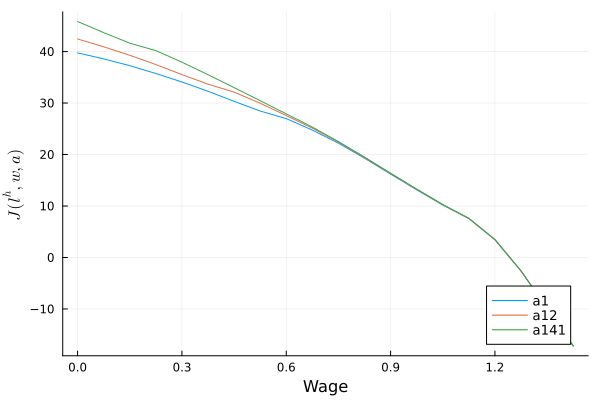}
  \subcaption{Firms Value with a High-Productivity Worker and Wage}
  \end{minipage}
  \end{tabular}
  \caption{Firms value.}
\end{figure}
\begin{figure}[H]
  \begin{tabular}{cc}
  \begin{minipage}[b]{0.45\textwidth}
  \centering
  \includegraphics[keepaspectratio, width=0.9\textwidth]{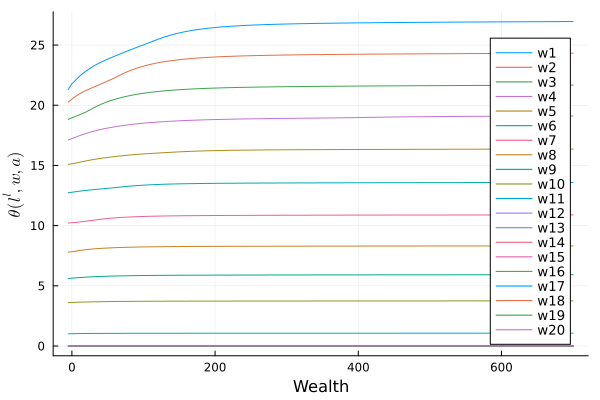}
  \subcaption{Market Tightness for a Low-Productivity Worker and Wealth}
  \end{minipage}&
  \begin{minipage}[b]{0.45\textwidth}
  \centering
  \includegraphics[keepaspectratio, width=0.9\textwidth]{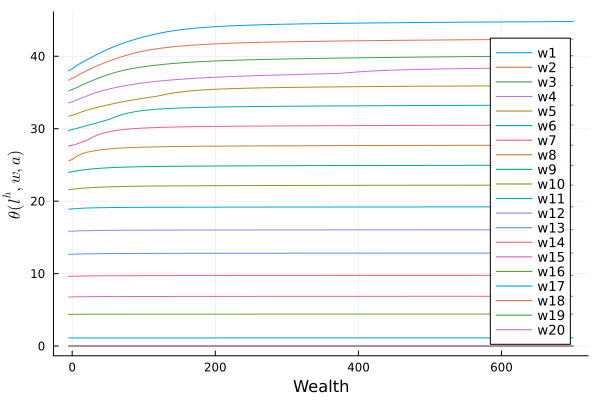}
  \subcaption{Market Tightness for a High-Productivity Worker and Wealth}
  \end{minipage}\\
  \begin{minipage}[b]{0.45\textwidth}
  \centering
  \includegraphics[keepaspectratio, width=0.9\textwidth]{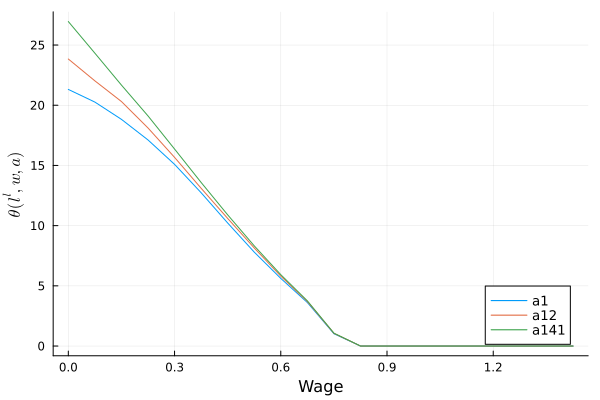}
  \subcaption{Market Tightness for a Low-Productivity Worker and Wage}
  \end{minipage}&
  \begin{minipage}[b]{0.45\textwidth}
  \centering
  \includegraphics[keepaspectratio, width=0.9\textwidth]{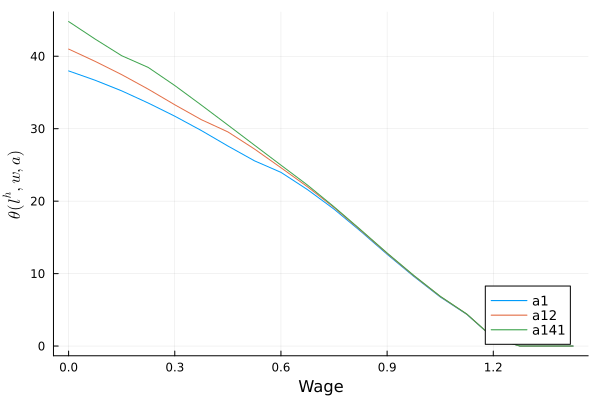}
  \subcaption{Market Tightness for a High-Productivity Worker and Wage}
  \end{minipage}
  \end{tabular}
  \caption{Market tightness.}
\end{figure}
\subsection{Minimum wages}
I solve three types of models, the baseline model, the model with the non-binding minimum wage, and the model with the binding minimum wage.
As mentioned in section 3.3, in the model with the non-binding minimum wage, I express it by setting $\underline{w}$ to 0.1.
Similarly, in the model with the binding minimum wage, I set $\underline{w}$ to 0.7.
In the remainder of this section, I compare the results of the baseline model to those of model with non-binding or binding minimum wage.
\par
\begin{table}[tp]
  \caption{Indicator values in the baseline and with minimum wages.}
  \label{T:Ivwm}
  \begin{center}\begin{small}
  \begin{tabularx}{\textwidth}{@{\extracolsep{\fill}}ll}
  \hline\hline\vphantom{\rule{0pt}{12pt}}%
  Indicator & Values (baseline $\rightarrow$ with the non-binding $\rightarrow$ with the binding) \\
  \hline\vphantom{\rule{0pt}{12pt}}%
  Unemployment rate of the low & 1.396\% $\rightarrow$ 1.396\% $\rightarrow$ 28.521\% \\
  Unemployment rate of the high & 1.178\% $\rightarrow$ 1.178\% $\rightarrow$ 1.600\% \\
  Job-to-job transition rate & 0.406\% $\rightarrow$ 0.406\% $\rightarrow$ 0.230\% \\
  \hline\hline
  \end{tabularx}
  \end{small}\end{center}
\end{table}
Table \ref{T:Ivwm} shows main resulting indicator values in the baseline model, the model with non-binding minimum wages, and the model with binding minimum wages.
We can see that non-binding minimum wages have few effects on the economy.
On the other hand, introducing binding minimum wages increases the unemployment rate of both low and high-productivity workers and decreases the job-to-job transition rate.
The magnitude of increase in the unemployment rate is higher among low-productivity workers.
\par
Figure 7 depicts the stationary distribution of the baseline model and that of the model with non-binding minimum wages.
Blue color bars and lines represent the stationary distribution in the baseline model.
Red color represents that in the model with the minimum wage.
However, in Figure 7, most of bars and lines are overlapped so that it seems that mixed color bars and lines are displayed.
Incorporating non-binding minimum wages has few effects on the stationary distribution.
\par
Figure 8 depicts the stationary distribution of baseline model and that of model with binding minimum wages.
From Panel (a), we can see that the binding minimum wage increases the unemployment rate and the number of employed workers with higher-paying jobs.
Panel (b) shows that the wealth distribution shifts upward with the introduction of the binding minimum wage.
\begin{figure}[H]
  \centering
  \begin{subfigure}[t]{\textwidth}
      \centering
      \includegraphics[width=1.00\textwidth]{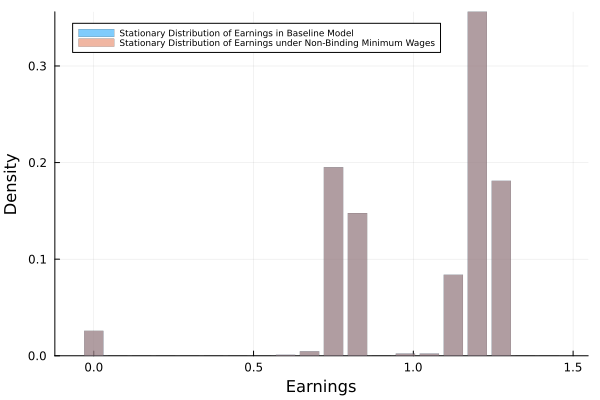}
      \caption{Stationary Distribution of Earnings of the Baseline and with Non-Binding Minimum Wages}
  \end{subfigure}
  \vspace{\baselineskip} 
  \begin{subfigure}[t]{\textwidth}
      \centering
      \includegraphics[width=1.00\textwidth]{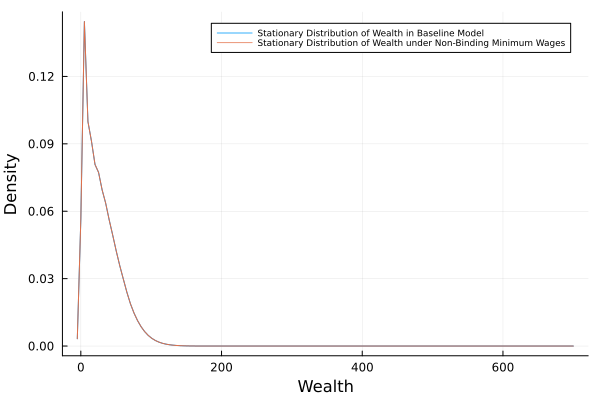}
      \caption{Stationary Distribution of Wealth of the Baseline and with Non-Binding Minimum Wages}
  \end{subfigure}
  \caption{Stationary distribution of the baseline and with non-binding minimum wages.}
\end{figure}
\begin{figure}[H]
  \centering
  \begin{subfigure}[t]{\textwidth}
      \centering
      \includegraphics[width=1.00\textwidth]{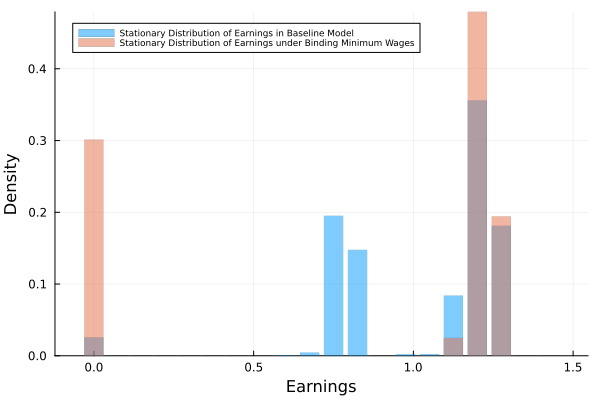}
      \caption{Stationary Distribution of Earnings of the Baseline and with Binding Minimum Wages}
  \end{subfigure}
  \vspace{\baselineskip} 
  \begin{subfigure}[t]{\textwidth}
      \centering
      \includegraphics[width=1.00\textwidth]{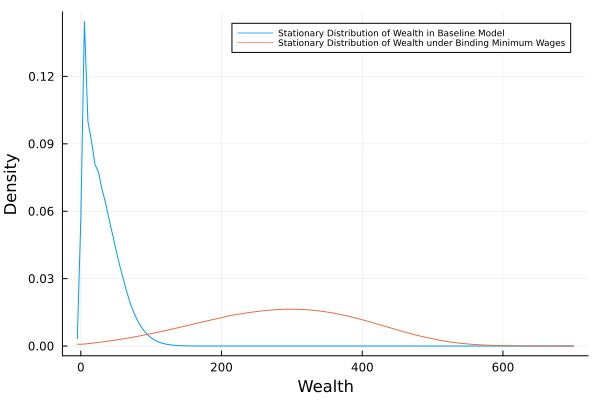}
      \caption{Stationary Distribution of Wealth of the Baseline and with Binding Minimum Wages}
  \end{subfigure}
  \caption{Stationary distribution of the baseline and with binding minimum wages.}
\end{figure}
\subsection{Discussion}
As traditionally argued, introducing binding minimum wages increases the unemployment rate.
According to the results of the policy experiment, low-productivity workers are strongly affected by the minimum wage.
This is because the binding minimum wage I incorporate eliminates jobs targeted only by low-productivity workers.
Even though jobs targeted by high-productivity workers are not eliminated, the unemployment rate of high-productivity workers also increases after incorporating the minimum wage.
Why are high-productivity workers also affected by the minimum wage?
Since workers' productivity fluctuates, then, some low-productivity unemployed workers affected by the minimum wage become high productivity unemployed workers.
This means that it is low-productivity workers who transmit the effect of the minimum wage on the employment of high-productivity workers.
\par
Additionally, by introducing the binding minimum wage, workers accumulate more wealth.
Due to the increased likelihood of unemployment, people begin saving more to smooth their consumption over time.
See Figure 9, which displays the value obtained by subtracting saving policy in the model with the binding minimum wage from that in the baseline model.
We can see that all values are less than or equal to zero meaning that workers tend to accumulate more wealth when the binding minimum wage is present.
\begin{figure}[H]
  \centering
  \begin{subfigure}[t]{\textwidth}
      \centering
      \includegraphics[width=0.66\textwidth]{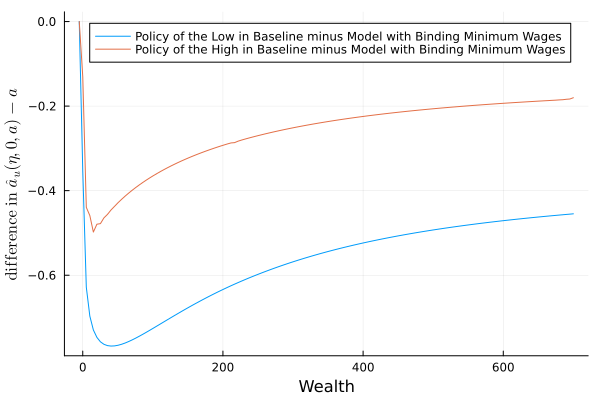}
      \caption{Differences in Saving Policy for the Unemployed between Both Models}
  \end{subfigure}
  \vspace{1em} 
  \begin{subfigure}[t]{\textwidth}
      \centering
      \includegraphics[width=0.66\textwidth]{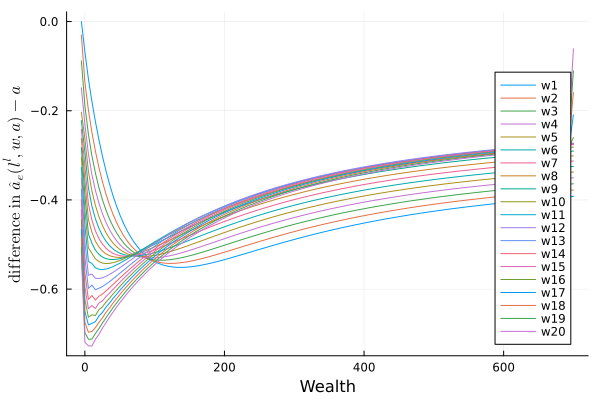}
      \caption{Differences in Saving Policy for the Low-Productivity Employed between Both Models}
  \end{subfigure}
  \vspace{\baselineskip} 
  \begin{subfigure}[t]{\textwidth}
      \centering
      \includegraphics[width=0.66\textwidth]{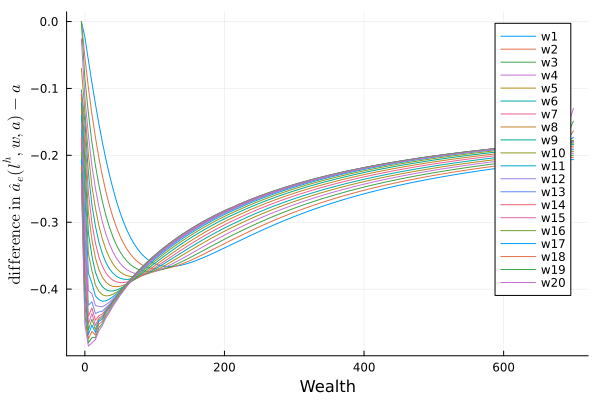}
      \caption{Differences in Saving Policy for the High-Productivity Employed between Both Models}
  \end{subfigure}
  \caption{Saving policy in the baseline minus that in the model with binding minimum wages.}
\end{figure}
\section{Conclusion}
I examine the effects of minimum wages on employment by use of a macroeconomic model incorporating productivity fluctuations and incomplete markets a la \citet{aiyagari1994uninsured},
directed search, and on-the-job search.
\par
I find that as traditionally argued, the binding minimum wage increases the unemployment rate.
However, not only low-productivity workers but also high-productivity workers experience an increase in the unemployment rate.
In addition, binding minimum wages drive households to accumulate more wealth.
\par
There are two possible extensions to this model.
First, we can discretize the AR(1) process into a Markov chain with more states.
Second, we can reincorporate the omitted assumption, that is, we can incorporate
the assumption that unemployment insurance is calculated based on the wage from the former position.
I think that this assumption makes the stationary distribution of earnings more realistic.
\bibliographystyle{aea}


\end{document}